\documentclass[9pt,twocolumn,twoside]{opticajnl}
\journal{opticajournal} 

\setboolean{shortarticle}{true}

\usepackage{lineno}
\usepackage{bm}

\newcommand{\be}{\begin{equation}}
\newcommand{\ee}{\end{equation}}
\newcommand{\bea}{\begin{eqnarray}}
\newcommand{\eea}{\end{eqnarray}}

\usepackage{color}

\title{A new direct method of continuous unwrapping phase
from a single interferogram}

\author[1,*]{V. BEREJNOV}
\author[2]{B. Y. RUBINSTEIN}

\affil[1]{6575 St. Charles Pl, Burnaby, V5H3W1, Canada}
\affil[2]{Stowers Institute for Medical Research, 1000 East 50th Street, Kansas City, MO 64110, USA}

\affil[*]{berejnov@gmail.com}

\begin{abstract}
A new method recovers the phase difference of interfering wavefronts from a pattern of interference fringes, avoiding the
phase discontinuity problem. The method relies on the numerical solution of one-dimensional first-order ordinary
differential equations. The solution of each equation allows to compute a corresponding cross section of the two-dimensional
phase, leading to the effective recovery of the phase surface. The unwrapping procedure can be performed
in each orthogonal direction.
\end{abstract}

\setboolean{displaycopyright}{false} 

\begin{document}

\maketitle

\smallskip
{\bf Introduction.}
Recovering the phase profile from a single interferogram has long
been of interest, despite its limitations in detecting large constant
phases and distinguishing between the concave and convex profile
shapes. However, this method remains valuable for applications
focused on relative phase changes and their models \cite{Aversano1997,Gokhale2004}, where
adjustments can be made to the constant phase and curvature sign.
For such applications, the trade-off between experimental setup
complexity and phase recovery typically favors the latter due to
challenges in obtaining even a basic single interferogram. The key
goal here is to recover a two-dimensional (2D) phase profile with a maximum number
of nodes for further modeling of the material characteristics \cite{Aversano1997,Gokhale2004}.
This process involves the following steps: obtaining the
interferogram, recovering the phase (adjusting curvature sign as
needed), converting it to optical path difference (OPD), and fitting
the OPD profile to describe material or geometrical parameters. The
accuracy of these parameters depends on the quality of fitting,
which explains the need for the high number of nodes in the OPD
profile. This requires access to all pixels available in the
interferogram, thus framing the problem of recovering the phase
per pixel.

Currently, two main approaches exist for recovering the phase
profile from a single interferogram. The first one uses the wrapped
discrete $2\pi$-phase pieces from the interferogram, leading to the $2\pi$
discontinuity problem. This issue is addressed by algorithmic
methods that unwrap the phase for each piece individually – the
{\it piecewise phase unwrapping} (PPU) approach \cite{Judge1994,Malacara2005}. The second
one, operating with the already unwrapped continuous phase – the
{\it continuous phase unwrapping} (CPU) approach, is represented by
a single method which involves simulating \linebreak

\begin{figure}[ht]
\centering
\includegraphics[width=7.5cm]{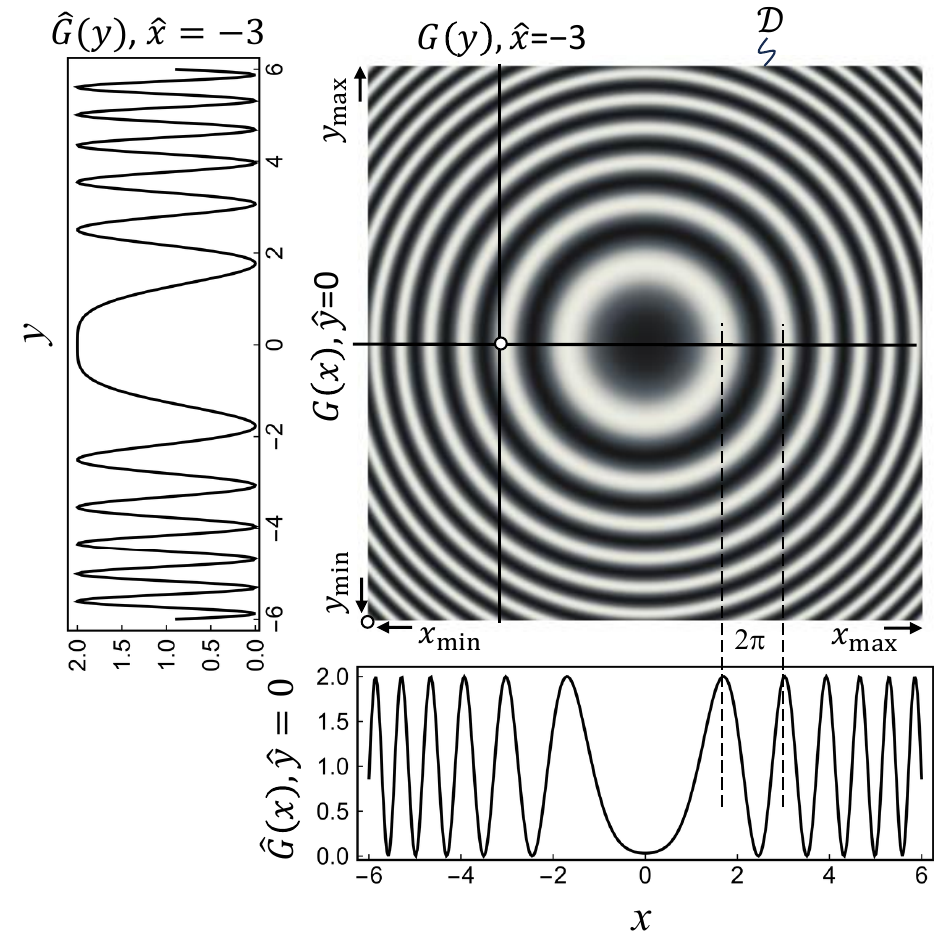}
\caption{The interferogram of the Example-1. Two 1D functions $\hat G(x)=G(x,0)$ 
and $\hat G(y)=G(-3,y)$
represent two cross-sections intersecting at $(-3,0)$
denoted by the white point in the interferogram. ${\cal D}$ represents the
interferogram boundary. The value $G_{00}=G(-6,-6)=2.0$ for computing
$\Theta_{00}$ is in the lower-left corner denoted by the white point.}
\label{Fig1}
\end{figure}

\noindent the entire 3D wavefront interference to represent the interferogram. This method employs
the 3D partial derivative transport intensity equation (TIE) for the unknown phase and solves it \cite{Pandey2016}.

We extend the CPU approach by developing an alternative to TIE
mathematical method that bypasses both the discontinuities
inherent in PPU methods and the need to simulate the 3D
wavefront for solving TIE. Our method employs 
1D first-order ordinary differential equations (ODEs) for the
interferogram function and the unknown continuous unwrapped
phase. The interferogram function is derived from a single
interferogram. Below, we present this new CPU method and
provide analytical and numerical examples for demonstration.

{\bf General definitions.} An interferogram is an image displaying a
pattern of bright (constructive) and dark (destructive) fringes
resulting from interference, see Fig.\ref{Fig1}. The fringe pattern is analyzed
within a Cartesian coordinate system and is characterized by the
gray values $G({\bf r})$ of the image pixels ${\bf r}=(x,y)$. These gray values
oscillate between minimum (black) and maximum (white) pixel
intensities in a region within a boundary ${\cal D}=\{(x,y):
x_{min}\le x \le x_{max},y_{min}\le y \le y_{max}\}$, see Fig.\ref{Fig1}. The function $G({\bf r})$
contains an information of a phase difference $\theta({\bf r})$ over the
interferogram.

Following reference \cite{Judge1994}, consider a general form of the function
$G$, suitable for the interference experiment where a single-pass
object beam in free space produces interference fringes of infinite
width, as illustrated in Fig.\ref{Fig1}. The fringe pattern to be analyzed is
\be
G = A + B \cos\theta,
\label{Eq1}
\ee
where the gray function $G$, the coefficients $A$ and $B$, and the phase $\theta$
are the functions of a particular point $(x,y)$ in the interferogram.
While $G$ is oscillating in space, the coefficients $A$ and $B$ are slowly
varying functions of spatial coordinates [3]. Depending on the
experimental conditions the coefficients $A$ and $B$ represent a
background illumination and an amplitude of the recorded light
modulation and we set them to constants.

{\bf CPU method.} Consider an arbitrary 2D interferogram obtained in
an experiment defined by Eq. (\ref{Eq1}). This interferogram represents a
gray function $G({\bf r})$ defined in the region bounded by ${\cal D}$. Introduce
the interferogram function $F({\bf r})$ – a relationship between the phase
difference $\theta({\bf r})$ and the gray function $G({\bf r})$, obtained from the
specified experimental conditions. Rewrite Eq. (\ref{Eq1}) to obtain $F$
\be
\cos\theta(x,y) = F(x,y) = (G(x,y)-A)/B,
\label{Eq2}
\ee
which also holds for the boundary ${\cal D}$. Differentiating Eq. (\ref{Eq2}) with
respect to $x$ and $y$ we find
\be
-\theta'_x \sin\theta = F'_x,
\quad
-\theta'_y \sin\theta = F'_y,
\label{Eq3}
\ee
where $f'_x = \partial f(x,y)/\partial x$ and $f'_y = \partial f(x,y)/\partial y$, 
respectively. Use Eqs.
(\ref{Eq2},\ref{Eq3}) in Pythagorean identity $\sin^2 \alpha + \cos^2\alpha = 1$ to eliminate the
trigonometric functions from consideration and obtain
\be
(\theta'_x)^2 = (F'_x)^2/(1-F^2),
\quad
(\theta'_y)^2 = (F'_y)^2/(1-F^2),
\label{Eq4}
\ee
that can be solved independently. The method of solution is
identical for both equations in Eq. (\ref{Eq4}), below we give a solution for
the $\theta'_x$ component only.

In Eq. (\ref{Eq4}) for the component $\theta'_x$, the coordinate $y$ is a parameter,
setting it to a specific value $y=\hat y$ turns $G,F,$ and $\theta$ into the functions
of the single variable $x$, namely $\hat G(x) = G(x,\hat y)$, $\hat F(x) = F(x,\hat y)$, 
and $\hat \theta(x) = \theta(x,\hat y)$ which are defined over the
interval $x_{min}\le x \le x_{max}$, where $x_{min}$ and $x_{max}$ values are taken from
the boundary of the interferogram, see Fig.\ref{Fig1}. 
A version of Eq. (\ref{Eq4})
written for the component $\theta_x$ with a single variable $x$ reads
\be
(\hat \theta'_x)^2 = (\hat F'_x)^2/(1-\hat F^2).
\label{Eq5}
\ee
Eq. (\ref{Eq5}) is 1D ODE of the first-order with the already unwrapped
phase. For representing the 2D profile of the phase, create $K$ 
functions $\hat F_k(x)$ taken from the interferogram pattern for a
set $Y=\{\hat y_k:1 \le k \le K\}$ of values $y_{min}\le \hat y_k \le y_{max}$.
In this case, Eq. (\ref{Eq5}) generates a series of phase profiles denoted as $\hat \theta_k(x) = \theta(x,\hat y_k)$.
Solution of this equation for the profiles $\hat \theta_k(x)$ requires corresponding initial
condition at the boundary $x=x_{min}$, specifically $\theta(x_{min},\hat y_k)$.

To solve Eq. (\ref{Eq5}) apply the relations
$\sqrt{f^2} = |f|$ where $|\cdot|$ denotes a modulus and $|f(x)| = \mbox{sgn} (f)\cdot f \ge 0$
and $\mbox{sgn} (\cdot)$ stands for the sign function returning $\pm 1$, and obtain
\be
\hat \theta(x) = \Theta_{0y} + \mbox{sgn}(\hat \theta'(x))
\int_{x_{min}}^x  \frac{|\hat F'(\xi)| d\xi}{\sqrt{1-\hat F^2(\xi)}}.
\label{Eq6}
\ee
Here $\Theta_{0y}$ denotes the boundary value $\theta(x_{min},\hat y_k)$, 
where the index $0$ indicates the minimal value, $x_{min}$, according to ${\cal D}$; 
$ \Theta_{0y}$ can be found from Eq. (\ref{Eq4}, right) solving it 
along the $y$-axis for $x= x_{min}$. The solution
similar to Eq. (\ref{Eq6}) reads
\be
\!\! \!\! \!\!\!\! \!\! 
\theta(x_{min},y) = \Theta_{00} + \mbox{sgn}(\theta'(x_{min},y))\!\!\!
\int_{y_{min}}^y \!\!  \frac{|F'(x_{min},\psi)| d\psi}{\sqrt{1-F^2(x_{min},\psi)}},
\label{Eq7}
\ee
where $\Theta_{00}$ denotes the boundary value $\theta(x_{min},y_{min})$
with the same convention for
indices. Eq. (\ref{Eq7}) defines the boundary conditions for Eq. (\ref{Eq6}) with
$y=\hat y_k$. The value of $\Theta_{00}$ can be determined from Eq. (\ref{Eq2})
\be
\cos\Theta_{00} = F_{00} = (G_{00}-A)/B,
\label{Eq8}
\ee
where $F_{00} =F(x_{min},y_{min}),$ 
with $G_{00}=G(x_{min},y_{min})$. 
The value of $G_{00}$ is taken from the lower-left
corner of the interferogram, as it is shown in Fig.\ref{Fig1}.  Eq. (\ref{Eq8}) has two
solutions for $\Theta_{00}$ in the interval $-\pi \le \Theta_{00} \le \pi$. While both
solutions satisfy the initial $G({\bf r})$ pattern, only one satisfies the
experimental conditions. Since $\Theta_{00}$ is a constant, it shifts the
2D phase profile $\theta(x,y)$ as a whole. Thus, either solution of Eq. (\ref{Eq8})
will not affect the general shape of the phase.

In Eqs. (\ref{Eq6},\ref{Eq7}), the integrand term at the first glance represents an
indeterminate $0/0$ for certain argument values. However, applying
the l’Hospital’s rule to the integrand quotient reveals that terms
contributing to a (possible) divergence cancel out. The function
$\mbox{sgn}(\hat \theta'(x))$ could be obtained once the extremum points $x_i$ of $\hat \theta(x)$
are found from the equation $\hat \theta'(x)=0$. As at these points both sides
of Eq. (\ref{Eq5}) vanish, the equation for the extrema $x_i$ reads
\be
(\hat F'_x)^2/(1-\hat F^2) = 0.
\label{Eq9}
\ee
Eq. (\ref{Eq9}) can be written in the finite difference form by using the
l’Hospital’s rule and solved numerically, same is applied for the
integrand in Eqs. (\ref{Eq6},\ref{Eq7}). For a sequence of $n$ roots $x_i, 1\le i \le n$, 
of Eq. (\ref{Eq9}), the whole interval should be divided into a sequence of $n+1$
segments $s_i:\{x_{i-1} \le x \le x_i\}$, $1\le i \le n+1$, where each segment is
characterized by a specific value of sign $\sigma_i=\mbox{sgn}(\hat \theta'(x))$ and $x_{n+1}=x_{max}$.
The sign alternates between the adjacent segments. Thus, the whole
sign sequence is determined by $\sigma_1$ in the first segment 
$s_1:\{x_{0}=x_{min} \le x \le x_1\}$ and it does not affect the generic shape of the unwrapped
phase.

During computation, the integral of Eq. (\ref{Eq6}) is applied per segment
$s_i$ while the argument $x \in s_i$, then the phase at the end point $x_{i-1}$ of
the preceding segment $\Theta_{i-1}=\theta(x_{i-1},\hat y)$ must be added. Thus, for
the interval $x_{i-1}\le x \le x_i$ the phase $\hat \theta_i(x)$ reads
\be
\hat \theta_i(x) = \Theta_{i-1} + \mbox{sgn}(\hat \theta'_i(x))
\int_{x_{i-1}}^x  \frac{|\hat F'(\xi)| d\xi}{\sqrt{1-\hat F^2(\xi)}}.
\label{Eq10}
\ee

For solution along the $y$-axis, replace the variable $x$ by $y$ in the
above 1D solution, while selecting $x=\hat x_m$. It outputs the stack of $M$
profiles $\hat \theta_m(y)=\theta(\hat x_{m},y)$ of the phase. These profiles require the
conditions $\Theta_{x0}=\theta(\hat x_{m},y_{min})$ along the $x$-coordinate, ultimately
leading to the same condition $\theta(x_{min},y_{min})=\Theta_{00}$ determined from
the interferogram, as done above.

\begin{figure}[ht]
\centering
\includegraphics[width=8.0cm]{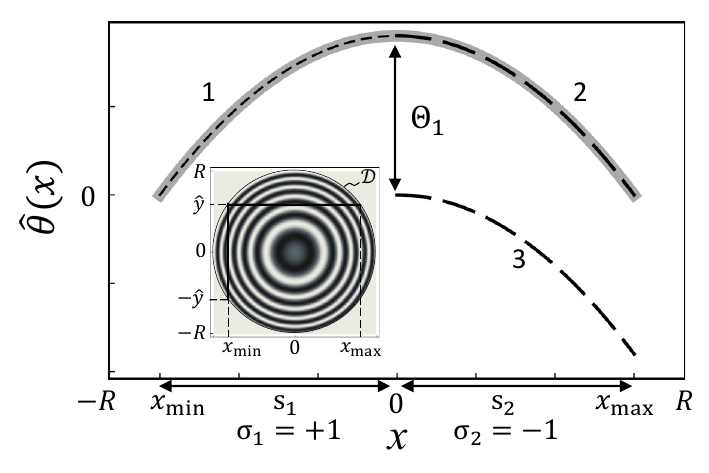}
\caption{Analytical example of 1D phase recovery. The black dashed curves $1$
and $2$ denote the phase $ \hat\theta_{1}(x)=\hat R^2-x^2$ and $ \hat\theta_{2}(x)=\Theta_1-x^2$
(with $\Theta_1=\hat R^2$) corresponding to the segments $s_1$ and $s_2$, 
where $x_{min}=-\hat R$ and $x_{max}=\hat R$, respectively.
The black dashed curve $3$ denotes the evaluation of the
integral in Eq. (\ref{Eq12}). The parabolic gray curve unites both dashed curves
denoting the resulting phase $\hat\theta(x)$. The insert shows the interferogram with
circular boundary ${\cal D}$ and the chord of interest $\hat y$.}
\label{Fig2}
\end{figure}

In general, for each point $(\hat x_{m},\hat y_{k})$ in the interferogram, our
method delivers two orthogonal components of the phase profiles
$\hat\theta_k(x)$ and $\hat\theta_m(y)$ by solving two independent 1D
ODEs effectively representing 2D phase surface. Once the 2D profile
of the phase difference $\theta$ is recovered, then the 2D optical path
difference (OPD) can be computed for the given refraction
coefficient of the media and the wavelength.

{\bf Analytical example.} To illustrate the method, consider
the ``seed'' phase difference defined as 2D, even, parabolic function:
$\theta(x,y)=R^2-x^2-y^2$ with the boundary 
${\cal D}=\{(x,y):R^2-x^2-y^2=0\}$, where $R$ is constant, Fig.\ref{Fig2} (insert). 
Consider the phase recovery task in 1D for an arbitrary chord $-R\le \hat y \le R$ limited
along the $x$-coordinate by  $x_{min}=-\hat R$ and $x_{max}=\hat R=\sqrt{R^2-\hat y^2}$.
Use Eq. (\ref{Eq1}) with same assumptions and obtain the gray
function $G(x)=A+B\cos(\hat R^2-x^2)$. The goal is to evaluate Eq. (\ref{Eq5}),
find the phase, and compare it with the ``seed''. The interferogram
function is given by $\hat F=\cos(\hat R^2-x^2)$ leading to $|\hat F'_x/\sqrt{1-\hat F^2}|=2|x|$.
Eq. (\ref{Eq9}) gives a condition for the roots $4x^2=0$. In the 
interval $-\hat R\le x \le \hat R$ there is a single root $x_1=0$, being an extremum point of the
phase function, so it produces only two constant sign segments. In
the first segment $s_1:\{-\hat R\le x \le 0\}$ the sign $\sigma_1=+1$, the integrand
of Eq. (\ref{Eq10}) is $2|x|$, and the phase $\hat\theta_{1}(x)$
for this interval reads
\be
\hat\theta_{1}(x) = \sigma_1
\int_{-\hat R}^x  2|\xi| d\xi = \hat R^2-x^2,
\label{Eq11}
\ee
note, $\Theta_{0y}=0$ as well as $\Theta_{00}=0$, 
because the phase vanishes at all
points of the boundary $\theta|_{\cal D}=0$ significantly simplifying
computation, Fig.\ref{Fig2}. In the second segment $s_2:\{0 \le x \le \hat R\}$ we have
$\sigma_2=-1$, and the phase for this interval  $\hat\theta_{2}(x)$
evaluates following
Eq. (\ref{Eq10}) to
\be
\hat\theta_{2}(x) = \Theta_1 + \sigma_2
\int_{0}^x  2|\xi| d\xi = \Theta_1-x^2,
\label{Eq12}
\ee
where $\Theta_1=\hat R^2$
is the phase value at the end point $x=0$ of the
preceding segment $s_1$ computed from Eq. (\ref{Eq11}). Uniting the
segments we obtain the final phase 
$\hat\theta(x) = \hat\theta_{1}(x)\cup \hat\theta_{2}(x)=\hat R^2-x^2$ 
for the entire interval $-\hat R\le x \le \hat R$. As $\hat R^2=R^2-\hat y^2$ and
$\hat y$ is arbitrarily selected we conclude that the final 2D phase $\theta(x,y)=
R^2-x^2-y^2$ coincides with the ``seed'' phase. Fig.\ref{Fig2} illustrates the
recovery procedure.
\begin{figure*}[ht]
\begin{center}
\begin{tabular}{ccc}
\includegraphics[width=5.7cm]{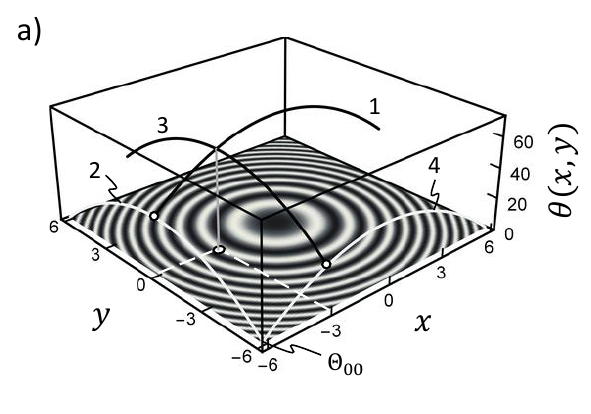} &
\includegraphics[width=5.7cm]{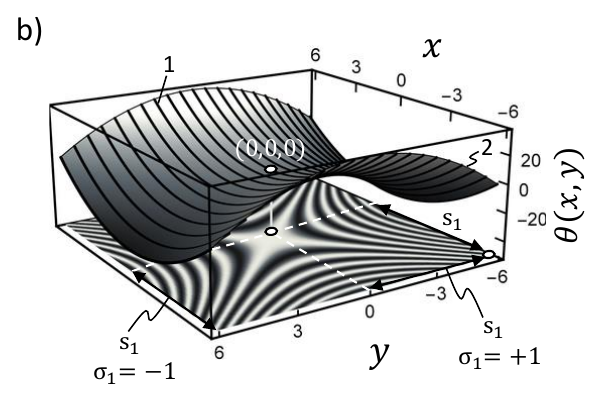} &
\includegraphics[width=5.7cm]{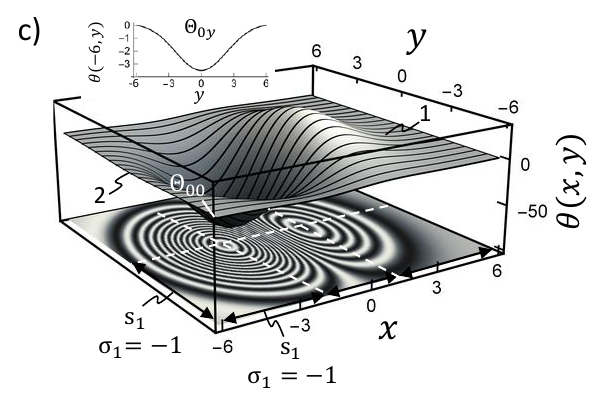} 
\end{tabular}
\end{center}

\caption{
Numerical recovery of phase from the interferogram 
having a) parabolic (Example-1), b) saddle (Example-2) and pit-and-hill (Example-3) type of the fringe pattern. 
a) Two orthogonal phase components are recovered. The black thick curves labeled $1$ and $3$ represent the numerical solutions $\theta(x,0)$ 
and $\theta(-3,y)$ of Eq. (\ref{Eq6}). The white vertical line is a projection of the
phase components’ intersection and the node of interest $(-3,0)$ in the
interferogram. White curves $2$ and $4$ are the boundary conditions $\Theta_{0y}$ and
$\Theta_{x0}$, respectively.
In b) and c) 2D shape is constructed by $21$
curves $\theta(x,y_b)$ recovered along $x$-direction, with typical profile denoted by $1$.
The curve of the boundary conditions $\Theta_{0y}$ for each $\theta(x,y_b)$ is marked by $2$.
Only the first segment $s_1$ and its sign $\sigma_1$ are shown for both $x$ and $y$ directions.
In c) $2$ denotes the curve of the boundary conditions $\Theta_{0y}$ shown in the insert for clarity.}
\label{Fig3}
\end{figure*}

{\bf Numerical examples.}
The 2D examples were
numerically tested using Mathematica 10.4, (Wolfram Research,
Inc.) using the following procedure: 
\begin{itemize}
\item
Define the seed phase $\theta_s$ in a rectangular region;
\item
Apply Eq. (\ref{Eq1}) and create an analytical $G(x,y)$ function with $A=B=1$,
leading to $G(x,y)=1+\cos\theta_s$ for all examples; 
\item
Create arrays $\{x_a, 1\le a \le 201\}$ of $201$ nodes and $\{y_b, 1\le b \le 21\}$ of $21$ nodes, 
and convert $G$ into the digital array $\{G_{a,b}\}$; 
compute the digital values of interferogram functions $\{F_{a,b}\}$; 
\item
Interpolate $\{F_{a,b}\}$
with splines along all $x$-nodes for each node $y_b$ creating an array of
the functions $F(x,y_b)$;
\item
Create $F(x_1,y)$ along $y$-nodes for the first $x$-node;
use for $F(x_1,y)$ Eq. (\ref{Eq9}) written for variable $y$ to find all roots $y_i$ (extrema), compute the segments $s_i$ and set their signs $\sigma_i$; 
then employ Eqs. (\ref{Eq10}) and (\ref{Eq7},\ref{Eq8}) and reconstruct the phase $\theta(x_1,y)$, this is a boundary
condition for reconstructing the phases $\theta(x,y_b)$ for given $y_b$ along $x$;
\item
For reconstructing each $\theta(x,y_b)$, set the $y$-node $y_b$ and for the
selected $F(x,y_b)$ solve Eq. (\ref{Eq9}) finding all roots $x_i$ (extrema), the
segments $s_i$ and set signs $\sigma_i$; 
then employ Eqs.(\ref{Eq10}) and (\ref{Eq7},\ref{Eq8}) and
obtain $\theta(x,y_b)$ taking into account the boundary condition $\theta(x_1,y_b)$;
then change the $y$-node and repeat.
\end{itemize}

{\bf Example-1} considers the parabolic seed phase $\theta_s(x,y)=72-x^2-y^2$ 
with the square boundary ${\cal D}=\{(x,y):-6\le x \le 6,-6\le y \le 6\}$ 
according to Fig.\ref{Fig3}a. The node $(-6,-6)$
provides $\Theta_{00}=0.0$ and two first-node arrays provide the phases at
the boundaries $\Theta_{0y}=\theta(-6,y)$ and $\Theta_{x0}=\theta(x,-6)$, respectively.
There is a single root $x_1=0.0$, making two segments $s_1:\{-6\le x \le 0\}$ 
and $s_2:\{0\le x \le 6\}$ with the selected signs $\sigma_1=+1$
and $\sigma_2=-1$. Fig.\ref{Fig3}a demonstrates a phase recovery along $x$ and $y$
directions corresponding to $\theta(-3,y)$ and $\theta(x,0)$ phase components
for the node $\theta(-3,0)$.

{\bf Example-2} illustrates the numerical recovery of the saddle type
of the seed phase $\theta_s(x,y)=x^2-y^2$  with the square boundary
${\cal D}=\{(x,y):-6\le x \le 6,-6\le y \le 6\}$. 
21 profiles $\theta(x,y_b)$ are
recovered producing an effective approximation of 2D phase
surface. The corresponding interferogram is presented in Fig.\ref{Fig3}b. The
boundary conditions for the $\theta(x_{min},y_{min})$ and the first $y$-node are
following: $\Theta_{00}=\theta(-6,-6)=0.0$ and $\Theta_{0y}=\theta(-6,y_b)$,
respectively. There is a single root $x_1=0.0$ along both directions,
providing for each two segments of integration $s_1$ and $s_2$, having
different signs $\sigma$. Along the $x$-axis $s_1$ has $\sigma_1=-1$, while along the $y$-axis
$s_1$ has $\sigma_1=+1$. Signs are selected by trials in order to match the
recovered phase with the initial seed phase.

{\bf Example-3} 
recovers the phase from the pit-and-hill 
interferogram, Fig.\ref{Fig3}c. The interferogram corresponds to
the seed phase $\theta_s(x,y)=1+50 x \exp(-(0.4x+0.3)^2-(0.3y)^2)$ with the
square boundary ${\cal D}=\{(x,y):-6\le x \le 6,-6\le y \le 6\}$.
The phase was recovered along $x$-axis $\theta(x,y_b)$ represented by 21
curves aligned with the 2D profile of $\theta_s$ in Fig.\ref{Fig3}c. 
The value $\Theta_{00}=\theta(-6,-6)$ is $0.857$; the boundary conditions for 
each $\theta(x,y_b)$ is $\Theta_{0y}=\theta(-6,y_b)$ shown in the Fig.\ref{Fig3}c insert. 
Solution of Eq. (\ref{Eq9}) gives
two roots along the $x$-axis $(-2.182,1.432)$
corresponding to the extrema and providing three segments $s_1,s_2$ and $s_3$
for integration in Eqs. (\ref{Eq6},\ref{Eq7},\ref{Eq10}); 
the sign for $s_1$ is selected $\sigma_1=-1$. There is one root $x_1=0.0$ along the $y$-axis providing two
segments $s_1,s_2$; the sign for $s_1$ is selected $\sigma_1=-1$. Both $\sigma_1$ are
selected by trials to match the initial phase seed (in practice the
experimental insights must be used). Recovering the phase along $y$-axis produces the $\theta(x_a,y)$ curves require changing the sign of $\sigma_1$
while progressing along $x_a$ nodes.

{\bf Discussion.} In summary, our method allows to recover 2D phase profile from
a single interferogram by analyzing fringe patterns of different
complexity. The user defines the interferogram function $F$, manages
the constant phase matching $-\pi \le \Theta_{00} \le \pi$, and initializes the sign
$\sigma_1$ for the first integration segment $s_1$. The method outputs the
recovered phase profile. To align with the interferogram and/or
experimental conditions, the user may need to change the sign of $\sigma_1$
and remove the numerically computed roots corresponding to
the inflection points along the integration path, addressing the
inherent insensitivity of the interferogram pattern to the phase
profile concavity or convexity. 

For complex patterns with varying
phase curvature, the method accurately
tracks the phase profile changes along the integration path taken
parallel to the line connecting the phase profile extrema in the
interferogram.

If the media where interference occurs have refractions and the
interfering wavefronts have reflections, the functions $G$ and $F$ differ
from those in Eqs. (\ref{Eq1},\ref{Eq2}). The 
preprint \cite{Berejnov2010} provides an example of
phase recovery using this new method in the more complex case
involving experimental interference in thin liquid films, which is
applicable to the contexts described in 
\cite{Aversano1997,Gokhale2004}.
The presented novel CPU method of continuous phase
unwrapping can be considered as a complementary one to the
existing PPU and CPU-TIE methods.

\begin{backmatter}
\bmsection{Disclosures} The authors declare no conflicts of interest.
\bmsection{Data availability} Data underlying the results of this paper is
presented in this paper.
\end{backmatter}

\bigskip


\ifthenelse{\equal{\journalref}{aop}}{%
\section*{Author Biographies}
\begingroup
\setlength\intextsep{0pt}
\begin{minipage}[t][6.3cm][t]{1.0\textwidth} 
  \begin{wrapfigure}{L}{0.25\textwidth}
    \includegraphics[width=0.25\textwidth]{john_smith.eps}
  \end{wrapfigure}
  \noindent
  {\bfseries John Smith} received his BSc (Mathematics) in 2000 from The University of Maryland. His research interests include lasers and optics.
\end{minipage}
\begin{minipage}{1.0\textwidth}
  \begin{wrapfigure}{L}{0.25\textwidth}
    \includegraphics[width=0.25\textwidth]{alice_smith.eps}
  \end{wrapfigure}
  \noindent
  {\bfseries Alice Smith} also received her BSc (Mathematics) in 2000 from The University of Maryland. Her research interests also include lasers and optics.
\end{minipage}
\endgroup
}{}

\end{document}